\documentclass[review]{elsarticle}
\usepackage{graphicx,times}
\usepackage{natbib}
\usepackage{amssymb,amsmath}
\usepackage{lineno,hyperref}
\usepackage{rotating}

\journal{New Astronomy}

\newcommand{\ser}{S\'ersic~}
\newcommand{\mh}{H_{\rm{F160W}}}

\begin{document}

\begin{frontmatter}

\title{Compact star forming galaxies as the progenitors of compact quiescent galaxies: clustering result}

\author[ad1]{Xiaozhi Lin}
\author[ad2,ad1]{Lulu Fan}
\ead{llfan@sdu.edu.cn}
\author[ad1]{Xu Kong}
\ead{xkong@ustc.edu.cn}
\author[ad3]{Guanwen Fang}

\address[ad1]{CAS Key Laboratory for Research in Galaxies and Cosmology, Department of Astronomy, University of Science and Technology of China, Hefei, Anhui 230026, China}
\address[ad2]{Shandong Provincial Key Lab of Optical Astronomy and Solar-Terrestrial Environment, Institute of Space Sciences, Shandong University,Weihai, 264209, China}
\address[ad3]{Institute for Astronomy and History of Science and Technology, Dali University, Dali 671003, China}

\begin{abstract}
We present a measurement of the spatial clustering of massive compact galaxies at $1.2\le z \le 3$ in CANDELS/3D-HST fields. We obtain the correlation length for compact quiescent galaxies (cQGs) at $z\sim1.6$ of $r_{0}=7.1_{-2.6}^{+2.3}\ h^{-1}Mpc$ and compact star forming galaxies (cSFGs) at $z\sim2.5$ of $r_{0}=7.7_{-2.9}^{+2.7}\ h^{-1}Mpc$ assuming a power-law slope $\gamma =1.8$. The characteristic dark matter halo masses $M_H$ of cQGs at $z\sim1.6$ and cSFGs at $z\sim2.5$ are $\sim7.1\times 10^{12}\ h^{-1} M_\odot$ and $\sim4.4\times10^{12}\ h^{-1} M_\odot$, respectively. Our clustering result suggests that cQGs at $z\sim1.6$ are possibly the progenitors of local luminous ETGs and the descendants of cSFGs and SMGs at $z>2$. Thus an evolutionary connection involving SMGs, cSFGs, QSOs, cQGs and local luminous ETGs has been indicated by our clustering result.
\end{abstract}



\end{frontmatter}


%
\section{Introduction}

Massive ($M_\star\ge 10^{10}M_\odot$), quiescent galaxies (QGs) at high redshift ($z\sim 2$) have been found to have $3-5$ times smaller effective radii than their local counterparts (e.g., Daddi et al. 2005; Trujillo et al. 2006; van der Wel et al.2008; van Dokkum et al. 2008; Damjanov et al. 2009; Newman et al. 2012; Szomoru et al. 2012; Zirm et al. 2012; Fan et al. 2013a, b). Since massive compact quiescent galaxies (thereafter cQGs) in the local Universe are rare (e.g., Poggianti et al. 2013), a significant structural evolution has been required. Therefore, there raised two questions: (1) how do these cQGs evolve into local luminous early-type galaxies (ETGs) with larger size? and (2) how did these cQGs form at higher redshift?

There are two physical mechanisms which have been proposed to explain the observed structural evolution of cQGs at $z\ge 1$. One is dissipationless  (dry) minor mergers  (Naab et al. 2009; Oser et al. 2012; Oogi et al. 2016).  The other is "puff-up" due to the gas mass loss by AGN  (Fan et al. 2008, 2010) or supernova feedback  (Damjanov et al. 2009). The recent evidence has shown the inside-out growth of massive cQGs at $z>2$, which indicates that dry minor mergers may be the key driver of structural evolution (Patel et al. 2013). However, whether dry minor mergers are sufficient for the size increase, especially at $z\ge 1.5$, is still under debate (Newman et al. 2012; Belli et al. 2014).

Possible mechanisms for the formation of cQGs include gas rich mergers (Hopkins et al. 2008), violent disk instability fed by cold stream, or both (Ceverino et al. 2015). Whatever mechanism governs the formation of cQGs, their precursors should be expected to experience a compact and
active phase: compact star forming galaxies (cSFGs) or compact starburst galaxies (i.e, sub-millimeter galaxies, SMGs). Barro et al. (2013) found a population of massive cSFGs at $z\sim2$. They proposed that cSFGs could be the progenitors of cQGs at lower redshift, suggested by the comparison of their masses, sizes, and number densities. Toft et al. (2014) showed that SMGs at $z>3$ are consistent with being the progenitors of $z\sim2$ cQGs by matching their formation redshifts and their distributions of sizes, stellar masses, and internal velocities. They suggested a direct evolutionary connection between SMGs, through compact quiescent galaxies to local ETGs. In this evolutionary scenario, star formation quenching has been proposed to be either due to gas exhaustion or quasar (QSO) feedback. The latter is essential in many models of the evolution of massive galaxies (e.g., Granato et al. 2004; Hopkins et al. 2010).

In this paper, we analyze the clustering properties of cQGs and cSFGs at $1.2\le z\le 3$ and compare them to other populations: high-$z$ QSOs, SMGs and local ETGs in order to investigate the possible connection between cQGs, cSFGs, SMGs, QSOs and local ETGs. All our data come from the CANDELS and 3D-HST programs (Grogin et al. 2011; Koekemoer et al. 2011; Skelton et al. 2014). The CANDELS/3D-HST programs have provided WFC3 and ACS images, spectroscopy and photometry covering $\approx 900$ arcmin$^{2}$ in five fields: AEGIS, COSMOS, GOODS-North, GOODS-South and the UDS. The large survey areas and the depth of the HST WFC3 camera enable us to make more accurate clustering measurement than in narrower, shallower fields. We emphasize that it is essential to use the high-resolution HST WFC3 imaging to investigate the compact structure of massive galaxies at high redshift. Throughout this paper, we adopt a flat cosmology (see Komatsu et al. 2011) with $\Omega_{M} = 0.3,\ \Omega_{\Lambda} = 0.7,\ H_{0} = 70\ km s^{-1} Mpc^{-1}$. We assume a normalisation for the matter power spectrum of $\sigma_{8} = 0.84$. All quoted uncertainties are 1 $\sigma$ (68\% confidence). All magnitudes are in the AB magnitude system.

\section{Data and Sample selection}

We select our massive compact galaxies at $1.2\le z \le 3$ from HST WFC3-selected photometric catalogs in the five CANDELS/3D-HST fields (Grogin et al. 2011; Koekemoer et al. 2011; Skelton et al. 2014)\footnote{http://candels.ucolick.org/}$^{,}$\footnote{http://3dhst.research.yale.edu/Data.php}. 
The five fields cover a total science area of 896 arcmin$^{2}$ after excluding the areas surrounding bright stars and field edge regions. For galaxies with $\mh < 23$ and having WFC3/G141 grism coverage, redshifts are measured using a modified version of the \texttt{EAZY} code (Brammer et al. 2013) from a combination of the $U-IRAC$ photometric data and the WFC3/G141 grism spectra. An accuracy of $0.003-0.005$ in $\Delta z /(1+z)$ can be reached by comparing to available spectroscopic redshifts. For the remaining galaxies, which are either faint or without grism spectra, photometric redshifts have been used instead. Probability distribution functions (PDFs) of redshift, or equivalently, the comoving line-of-sight distance $\chi$ is derived by minimising the chi-square in the photometric analysis using \texttt{EAZY} (Brammer et al. 2008). PDF for each galaxy is defined as $f(\chi)$, such that $\int f(\chi)d\chi=1$. The galaxy physical properties, such as stellar masses ($M_{\star}$), luminosity-weighted ages and rest-frame colors, are derived using \texttt{FAST} (Kriek et al. 2009), adopting Bruzual \& Charlot (2003) models assuming a Chabrier (2003) IMF, solar metallicity, exponentially declining star formation histories (SFHs) and Calzetti extinction law (Calzetti et al. 2000). For the measurement of effective radius $r_e$, we use the result in van der Wel et al. (2012)\footnote{http://www.mpia.de/homes/vdwel/candels.html}, which is based on best-fitting of \ser model.

In Figure 1, we show our selection criteria of massive compact galaxies at $1.2\le z\le 3$ on mass-size plane. We select compact galaxies at $1.2\le z \le 2$ by using the same criterion as presented by Barro et al. (2013) and the lower mass limit of $1.0\times10^{10} M_{\odot}$ (dotted line):
\begin{equation}
log(\Sigma_{1.5})\equiv log(M_{\star}/r_{e}^{1.5})  > 10.3\ M_{\odot}\cdot kpc^{-1.5}
\end{equation}

Similarly, we select compact galaxies at $2< z \le 3$ by using the same criterion as that in Barro et al. (2014) (dashed line):
\begin{equation}
log(\Sigma_{1.5})\equiv log(M_{\star}/r_{e}^{1.5})>10.45\ M_{\odot}\cdot kpc^{-1.5}
\end{equation}
Here we also impose a lower mass limit of $1.0\times10^{10} M_\odot$.

The rest-frame UVJ color diagram has been used to classify our compact sample into two classes: cSFGs and cQGs. This method has weak dependence on dust extinction and works well up to redshift 3 (e.g., Wuyts et al. 2007; Williams et al. 2009).


For the cross-correlation analysis, we also need two comparison galaxy samples at $1.2\le z \le 2$ and $2< z \le 3$ in the same fields. We take  $\approx14000$  and $\approx13000$ galaxies with mass range $10^9 M_\odot \le M_\star \le 10^{10}M_\odot$ within the redshift range $1.2\le z \le 2$ and $2< z \le 3$, respectively.


\section{Clustering analysis}

Our clustering analysis is identical to the QSO-galaxy and SMGs-galaxy cross-correlation study presented in Hickox et al. (2011) and Hickox et al. (2012). Here we summarize some key details.

The two-point correlation function $\xi(r)$ is defined by :
\begin{equation}
dP=n[1+\xi(r)]dV
\end{equation}
where $dP$ is the probability above Poisson of finding a galaxy in a volume element $dV$ at a physical separation $r$ from another randomly chosen galaxy, and $n$ is the mean space density.
In the linear halo-halo regime, the correlation function is well-described by a power law
\begin{equation}
\xi(r)=(r/r_{0})^{-\gamma}
\end{equation}
where $r_{0}$ is  the real-space correlation length and  $\gamma$ has a typical value of 1.8 (e.g. Peebles 1980).

By integrating  $\xi(r)$, we can obtain the projected correlation function $\omega_{p}(R)$:
\begin{equation}
\omega_{p}(R)=2\int_{0}^{\pi_{max}}\xi(R,\pi)d\pi
\end{equation}
where $R$ and $\pi$ are the radial and perpendicular projected comoving distances between the two galaxies in the view of the observer.
By averaging over all line-of-sight peculiar velocities, $\omega_{p}(R)$ can be re-written as:
\begin{equation}
\omega_{p}(R)=R(\frac{r_{0}}{R})^{\gamma}\frac{\Gamma(1/2)\Gamma((\gamma-1)/2)}{\Gamma(\gamma/2)}
\end{equation}

By weighing the PDFs of comparison galaxies overlapped with the redshift distribution of compact galaxy samples in matched pairs, we derive the real-space projected cross-correlation function using the method in Myers et al. (2009).
\begin{equation}
\omega_{p}(R)=N_{R}N_{C}\sum_{i,j}c_{i,j}\frac{D_{C}D_{G}(R)}{D_{C}R_{G}(R)}-\sum_{i,j}c_{i,j}
\end{equation}
where  $c_{i,j}=f_{i,j}/\sum_{i,j}f_{i,j}^{2}$ and  $f_{i,j}$ is defined as the average value of the radial PDF $f(\chi)$ for each comparison galaxy $i$, in a comoving distance window ($100 h^{-1}\ Mpc$) around each compact galaxy $j$. $R$ is the projected comoving distance from each galaxy in our compact galaxy sample to that in the comparison galaxy sample or random sample. For a given angular separation $\theta$ and radial comoving distance $\chi_{*}$ to the compact galaxy, $R$, $\theta$ and $\chi_{*}$ satisfy the relationship, $R=\chi_{*}\theta$. $D_{C}D_{G}$ and $D_{C}R_{G}$ are the numbers of compact-comparison galaxy pairs and compact-random galaxy pairs in each bin of $R$. $N_{C}$ and $N_{R}$ are the total numbers of compact and random galaxies, respectively. We calculate the pair count for each galaxy in the compact sample individually, and in this case, $N_{C}=1$.

To account for the fact that the transverse comoving distance (and thus the conversion between angle and projected physical distance) changes with redshift, for our cross-correlation analysis we divide our cQG and cSFG samples into small redshift bins with $\delta z=0.2$. The calculations of $D_{C}D_{G}$ and $D_{C}R_{G}$ are performed to derive the $\omega_{p}(R)$ values in these small redshift bins, ensuring that the comoving distance variations are small enough that each bin in R corresponds to a comparable range in angular separation. To minimize the effects of shot noise, we then average the $\omega_{p}(R)$ values in the different redshift bins, weighted by the relative sample size in each redshift bin, to derive the mean $\omega_{p}(R)$ values for cQG and cSFG samples. The calculations of $D_{C}D_{G}$ and $D_{C}R_{G}$ are performed to derive the $\omega_{p}(R)$ values in these small redshift bins. Then we average the $\omega_{p}(R)$ values, weighted by the relative sample size, to derive the mean $\omega_{p}(R)$ values for cQG and cSFG samples.
 {We derive the projected cross-correlation functions of five cQG subsamples at $1\le z\le 2$ (see the inset plot of Figure 3a). We can find that the projected cross-correlation function of cQG subsample at $1\le z< 1.2$ shows a large discrepancy at $R<2~h^{-1}Mpc$ and $R>8~h^{-1}Mpc$, and obvious peaks and troughs at $R\sim 2-4~h^{-1}Mpc$, compared to those of four cQG subsamples at higher redshift bins.
The former may be due to shot noise and the limited areas of the survey fields. 
Taking $R=10 h^{-1}\ Mpc$ as an example, the corresponding angular separation is about 13.9 arcmin at $z=1.1$, which is close to the boundary of the survey region (each survey region is about 180 arcmin$^{2}$ in size). The latter are likely due to the impact of the mask regions. We do a simple test by generating a random galaxy sample without removing mask regions and cross-correlating it with the $1\le z< 1.2$ cQG subsample and comparison sample. The result is shown as the dot-dashed line in the inset plot of Figure 3a. We can find that the peaks and troughs at $R\sim 2-4~h^{-1}Mpc$ disappear.
We also find that the comparison galaxies around $1\le z< 1.2$ have a sharp PDF of redshifts, which makes the cross-correlation function have a large scatter.
In order to improve the accuracy of the clustering analysis, we decide to discard the cQG subsample at $1\le z< 1.2$ and adopt the redshift range $1.2\le z \le 2$ for further analysis.}

We use the HALOFIT code of Smith et al. (2003) to generate the nonlinear-dimensionless power spectrum of the dark matter (DM) assuming the standard cosmology. The fourier transform of the power spectrum gives us the real-space correlation function of the DM. By integrating it to $\pi_{max}=100 h^{-1}\ Mpc$ (see Equation 5), we derive the projected correlation function $\omega_{p}(R,z)$ for the DM. Then we average the $\omega_{p}(R,z)$ over the redshift distribution of the samples, weighted by their overlap with the PDFs of the comparison galaxy samples, to derive the mean $\omega_{p}(R)$ for the DM at $1.2\le z\le3$ (see Figure 3). We perform a Monte Carlo integration of Equation(A6) of Myers et al. (2007) to obtain $\omega(\theta)$ for the DM.

We measure the angular autocorrelation function of comparison galaxies using the Landy \& Szalay et al. (1993) estimator:
\begin{equation}
\omega(\theta)=\frac{DD-2DR+RR}{RR}
\end{equation}
where DD, DR and RR are the number of data-data, data-random and random-random galaxy pairs at the separation $\theta$.

The integral constraint is defined as:
 \begin{equation}
 \omega_{\Omega}=\frac{1}{\Omega^{2}}\iint\omega(\theta_{12})d\Omega_{1}d\Omega_{2},
 \end{equation}
 which can significantly affect the clustering amplitude when the field is limited in size. We include the integral constraint for the calculation of both angular autocorrelation function of comparison galaxies and cross-correlation function of each galaxy sample. We correct the angular autocorrelation functions of comparison galaxies by their integral constraint. We determine the value of the cross-correlation functions of galaxy samples at $0.6-1h^{-1} Mpc$ and $1-1.2h^{-1} Mpc$ for $z\sim1.6$ and $z\sim2.5$, respectively, corresponding to approximate $1'$ at each redshift bin. We average this value and multiply it by the fraction of the integral constraint to $\omega(1')$ for comparison galaxies to derive the correction for the cross-correlation function. Accounting for the integral constraint, the clustering amplitudes of $1\le z\le2$ cQGs and $2<z\le3$ cSFGs will increase by $0.23\times \omega_{p}(0.6-1h^{-1} Mpc)$ and $0.40\times \omega_{p}(1-1.2h^{-1} Mpc)$, respectively.

 As all our samples are above the mass-completeness limit, their redshift distribution are similar at each redshift bin (Figure 2). The mean value and variance of redshift distributions differ by $<1\%$ for cSFGs/cQGs and their comparison samples.

Uncertainties in the angular autocorrelation function are derived using the covariance matrix calculated as in Brown et al. (2008). We use the bootstrap method to determine uncertainties in the cross-correlation function, which is explained in Hickox et al. (2011). Briefly speaking, we divide the survey volume into $N=8$ subvolumes, and randomly draw 3N subvolumes. The calculation of cross-correlation is repeated in these subvolumes. For simplicity, we calculate the variance between the result of different bootstrap samples.

We fit the observed $\omega_{p}(R)$ of the compact-comparison galaxies cross-correlation on scales { $1-10 h^{-1}Mpc$ } and transfer it to a simple linear scaling of angular correlation function $\omega(\theta)$, using Equation (A16) of Hickox et al. (2011). The best-fit linear scaling of $\omega({\theta})$ of compact and comparison galaxies to that of DM corresponds to $b_{C}b_{G}$, which is the product of the linear bias of the compact and comparison galaxies.

We obtain $b_{G}^{2}$ for the comparison galaxies from their angular autocorrelation in a similar manner to that has been applied to the compact-comparison galaxies cross-correlation. Thus we derive the bias of compact galaxies $b_{C}$, by combining with the cross-correlation measurement. Finally, we convert $b_{C}$ and $b_{G}$ to halo mass $M_{H}$ for each galaxy population using the prescription of Sheth et al. (2001).

\section{Result and Discussion}

We use a power law with a slope fixed to $\gamma=1.8$ to fit the projected cross-correlation functions of the compact-comparison samples (Figure 3a). And also, we fit the angular correlation function $\omega(\theta)$ for two comparison galaxy samples with a slope $\delta=0.8$ (Figure 3b). The bump at $4'\sim5'$ for comparison galaxies at $z\sim2.5$ is mainly due to the impact of the mask regions. As we are performing clustering analysis in 5 small regions, the areas of the mask regions and the size of each field will affect the result on larger scale.  For the purpose of comparison, we re-calculate the angular correlation function without removing mask regions inside each field. The result is shown in the inset plot of Figure 3b, illustrating that the bump disappears in this case. However, we do not use this result for further calculations in this paper, because the real space clustering amplitude will be artificially decreased due to the inclusion of the unreliable data in the regions that were originally masked. From the best-fit parameters of the cross-correlation for the compact and comparison galaxies and the autocorrelation of comparison galaxies, we derive $b_{C} = 2.74_{-0.84}^{+0.86}$ for cQGs at $1.2\le z\le2$, and $b_{C} = 3.72_{-1.19}^{+1.23}$ for cSFGs at $2< z\le3$. Converting these to DM halo masses using the prescription of Sheth et al. (2001), we obtain $log(M_{H}\ [h^{-1}M_{\odot}])=12.85_{-0.67}^{+0.41}$ for cQGs at $1.2\le z\le2$ and $log(M_{H}\ [h^{-1}M_{\odot}])= 12.64_{-0.67}^{+0.42}$ for cSFGs at $2< z\le3$. The corresponding DM halo masses for the comparison galaxies are $log(M_{H}\ [h^{-1}M_{\odot}])=11.60_{-0.37}^{+0.27}$ and $11.57_{-0.54}^{+0.23}$ at $z\sim1.6$ and $z\sim2.5$, respectively.

We obtain the correlation length $r_{0}$ by estimating the autocorrelation of compact galaxy samples from the cross-correlation through the relationship, $\xi_{CC} = \xi_{CG}^{2}/\xi_{GG}$ (Coil et al. 2009), with $r_{0}=7.1_{-2.6}^{+2.3} {h^{-1}Mpc}$ for cQGs at $1.2\le z\le2$ and $r_{0}=7.7_{-2.9}^{+2.7} {h^{-1}Mpc}$ for cSFGs at $2< z\le3$. Using the same process we derive the spatial clustering of compact galaxy samples. We summarize our results in Table 1.

As a comparison, we also fit the correlation functions with the slope $\delta$ as a free parameter. The slope $\delta$ of angular autocorrelation function of the comparison galaxies at $z\sim1.6$ and $z\sim2.5$ will be $0.87\pm0.13$ and $0.85\pm0.18$, respectively. Similarly, the cross-correlation function has been fitted with a variable slope $\gamma$. In this case, the derived correlation length $r_{0}$ of cQGs and cSFGs at $z\sim1.6$ and $z\sim2.5$ have the value of $5.27\pm0.76 h^{-1}Mpc$ and $5.60\pm0.92 h^{-1}Mpc$ with $\gamma=1.81\pm0.37$ and $1.70\pm0.38$ respectively, compared to $5.26_{-1.23}^{+1.14} h^{-1}Mpc$ and $5.69_{-1.70}^{+1.02} h^{-1}Mpc$ with $\gamma$ fixed to 1.8.



For a given DM halo mass $M_H$ and redshift $z$, we compute the corresponding correlation length $r_0(M_H,z)$ by fitting a power-law with $\gamma=1.8$ to the DM correlation function. In this way, we can determine the evolution of $r_0$ with redshift for given DM halo mass (dotted lines in Figure 4). For DM haloes hosting cQGs at $z\sim1.6$, we estimate their median mass growth with redshift $M_H(z)$, where $log M_H(z=1.6)$ is $12.85\ h^{-1}M_{\odot}$, using the median growth rate described by Equation 2 in Fakhouri et al. 2010 (see Figure 4). The expected evolution in $r_0$ for DM haloes hosting cQGs at $z\sim1.6$ can therefore be calculated (red dashed line in Figure 4).  The observed $r_0$ of cQGs at $z\sim1.6$ shows a weak evolution with redshift, changing from 7.3 ${h^{-1}Mpc}$ at $z\sim2.5$ to 7.0 ${h^{-1}Mpc}$ at $z\sim0$. The expected $r_0$ ($7.3 {h^{-1}Mpc} $) is consistent with the observed $r_0$ of cSFGs at $z\sim2.5$, $r_{0}=7.7_{-2.9}^{+2.7} {h^{-1}Mpc}$. As shown in Figure 5, the evolution of DM halo mass with redshift indicates that the typical progenitors of cQGs at $z\sim1.6$ would have halo mass $logM_H\sim12.6 h^{-1}M_{\odot}$ at $z\sim2.5$, which is consistent with halo mass of cSFGs at $z\sim2.5$.  Both results confirm the previous arguments by Barro et al. (2013,2014) in which cSFGs at higher redshift are possible progenitors of cQGs at lower redshift.

We also compare the clustering amplitudes of our massive compact galaxy samples with other galaxy populations over a range of redshift. The correlation lengths of SMGs (Hickox et al. 2012) and QSOs (Ross et al. 2009) at $z\sim2$  have been over-plotted in Figure 4. Similar to cSFGs, the halo mass and $r_0$ for SMGs match well with the evolution of halo mass and $r_0$ of cQGs at $z\sim1.6$ (See Figure 3 and 4). This result confirms the argument of SMGs as progenitors of cQGs (Toft et al. 2014). Recent studies using ALMA imaging have revealed the compact structure in SMGs (e.g., Ikarashi et al. 2015; Chen et al. 2015), which suggest that SMGs have similar density structure as cQGs and therefore could be the progenitors of cQGs. Our results show that they have comparable large scale clustering for three different population at $z\sim2$ : cSFGs, SMGs and QSOs. This suggests that cSFGs at high redshift may be, like SMGs and QSOs, a transient population of local luminous ETGs at their early evolutionary stage (e.g., Fang et al. 2015).

The descendants of cQGs at $z\sim1.6$ will likely be the luminous ETGs ($\sim$ 1 $L^{\star}$) in the local Universe according to comparison of their large-scale clustering (Figure 4 and Figure 5). An evolutionary connection has therefore been suggested that cSFGs and SMGs evolve into cQGs by star formation quenching, either due to gas exhaustion or QSOs feedback, and finally evolve into local luminous ETGs (Sanders \& Mirabel 1996; Granato et al. 2004; Hopkins et al. 2008; Alexander \& Hickox 2012).

We also estimate the lifetimes of cQGs at $z\sim1.6$, cSFGs at $z\sim2.5$  and SMGs at $z\sim2.0$.
The lifetime of a given galaxy sample can be expressed as:
\begin{equation}
t_{sample}=\Delta t\frac{n_{sample}}{n_{halo}}
\end{equation}
where $\Delta t$ is the time interval between the redshift range, $n_{sample}$ and $n_{halo}$ are the space densities of the corresponding galaxy sample and DM haloes. We use the halo mass function in Sheth et al. (2001) to derive the space densities of DM haloes by assuming a constant density growing rate. The space densities of haloes with $log(M_{H}\ [h^{-1}M_{\odot}])=12.85_{-0.67}^{+0.41}$ at $z\sim1.6$ and $log(M_{H}\ [h^{-1}M_{\odot}])= 12.64_{-0.67}^{+0.42}$ at $z\sim2.5$ are $d n_{halo}/d ln M_H = 2.7_{-2.3}^{+14.6}\times 10^{-4}Mpc^{-3}$ and $d n_{halo}/d ln M_H = 5.3_{-4.1}^{+22.1}\times 10^{-4}Mpc^{-3}$, respectively. By using a different halo mass function (e.g., Tinker et al. 2008), the space densities of haloes are $d n_{halo}/d ln M_H = 2.6_{-2.2}^{+13.6}\times 10^{-4}Mpc^{-3}$ and $d n_{halo}/d ln M_H = 4.8_{-3.8}^{+19.4}\times 10^{-4}Mpc^{-3}$. The space densities of $z\sim1.6$ cQGs and $z\sim2.5$ cSFGs are $3\times10^{-4} Mpc^{-3}$ and $9\times10^{-5} Mpc^{-3}$, respectively. As a comparison,  the space density of SMGs at $z\sim2.0$ is $2\times10^{-5} Mpc^{-3}$ (Hickox et al. 2012). The corresponding lifetimes of cQGs at $z\sim1.6$ and cSFGs at $z\sim2.5$ are 4.3Gyr and 315Myr. By using the halo mass function of Tinker et al. 2008, the results are 4.5Gyr and 351Myr, respectively. The large lifetime for cQGs at $z\sim1.6$ is mainly due to the constant density growing rate. If we allow the density growing rate to vary with time following the halo mass evolution track of cQGs in Figure 5, the corresponding lifetimes for cQGs at $z\sim1.6$ will be scaled down to 1.4Gyr and 1.5Gyr using the halo mass functions in Sheth et al. (2001) and Tinker et al. (2008), respectively. The lifetimes of cSFGs at $z\sim2.5$ and SMGs at $z\sim2$ ($\sim100Myr$) are similarly short, suggesting that cSFGs may lie at a transient phase of
the evolution of massive galaxies.

\section{Conclusion}

In this paper, we measure the cross-correlation between massive compact galaxies and comparison galaxies at $1.2\le z\le 3$ in CANDELS/3D-HST fields. We obtain the correlation length for cQGs at $z\sim1.6$ of $r_{0}=7.1_{-2.6}^{+2.3} h^{-1}Mpc$ and cSFGs at $z\sim2.5$ of $r_{0}=7.7_{-2.9}^{+2.7} h^{-1}Mpc$. The characteristic DM halo masses of cQGs at $z\sim1.6$ and cSFGs at $z\sim2.5$ are $\sim7.1\times 10^{12}\ h^{-1} M_\odot$ and $\sim4.4\times10^{12}\ h^{-1} M_\odot$, respectively. The observed clustering suggests that cSFGs and SMGs at $z>2$ could be the progenitors of cQGs at $z<2$. We estimate both the co-moving space densities and the corresponding lifetimes of cSFGs/cQGs and find that cSFGs have similarly short lifetime as SMGs. Our clustering results support such an evolutionary sequence involving compact starbursts (SMGs or cSFGs), cQGs and ETGs (see also Toft et al. 2014).

\section*{Acknowledgements}

We thank the referee for the careful reading and the valuable comments that helped improving our paper.
This work is based on observations taken by the CANDELS/3D-HST Program with the NASA/ESA HST. This work is supported by the Strategic Priority Research Program "The Emergence of Cosmological Structures" of the Chinese Academy of Sciences (No. XDB09000000), the National Basic Research Program of China (973 Program)(2015CB857004), and the National Natural Science Foundation of China (NSFC, Nos. 11673004,11203023, 11303002,11225315, 1320101002, 11433005 and 11421303). LF acknowledges the Qilu Young Researcher Project of Shandong University and the Knut and Alice Wallenberg Foundation for support. We thank Dr. Huiyuan Wang and Dr. Lixin Wang for valuable discussion.

\clearpage


\begin{figure}
\begin{center}
   \includegraphics[width=0.7\columnwidth,angle=90.0]{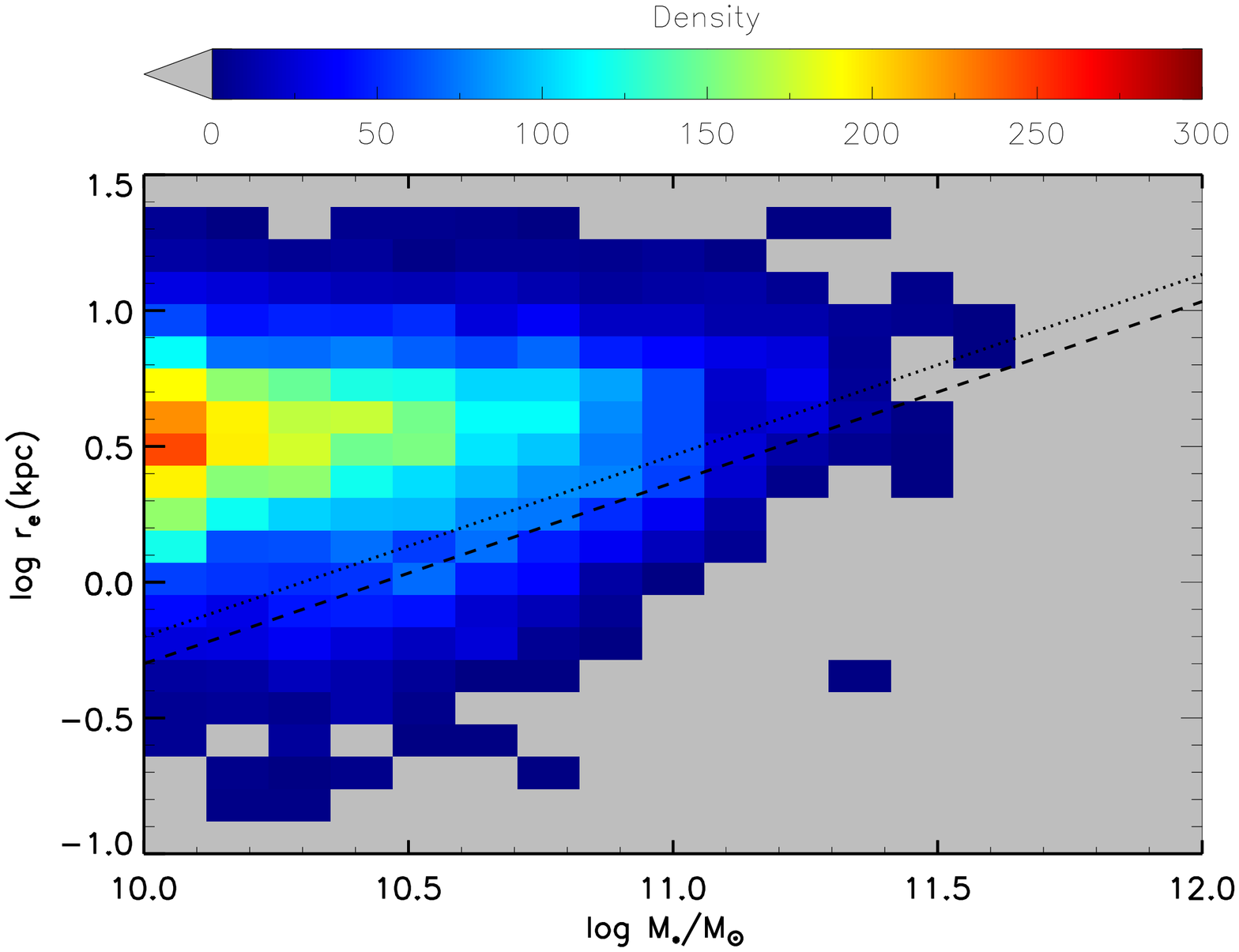}
   \caption{The density distribution of massive galaxies at $1.2\le z \le 3$ on the mass-size plane in CANDELS/3D-HST fields. The dotted and dashed lines mark the selection criteria of compact galaxies at  $1.2\le z \le 2$ and $2< z \le 3$, respectively. The top color bar shows the galaxy number density.}
\end{center}
   \label{Fig1}
\end{figure}

\begin{figure}
\begin{center}
   \includegraphics[width=0.7\columnwidth]{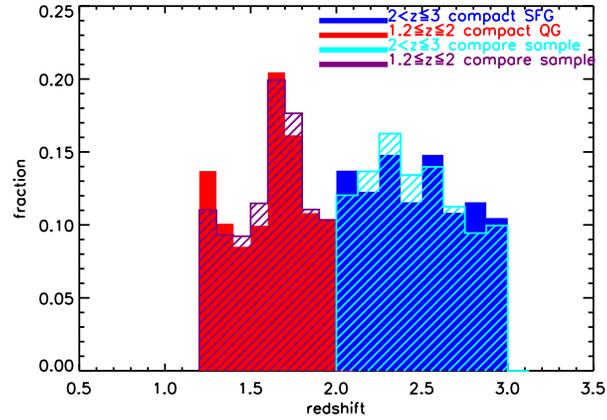}
   \caption{The redshift distributions of different galaxy samples at two redshift bins.}
\end{center}
   \label{Fig2}
\end{figure}

\begin{figure*}
\begin{center}
   \includegraphics[ width=0.495\textwidth]{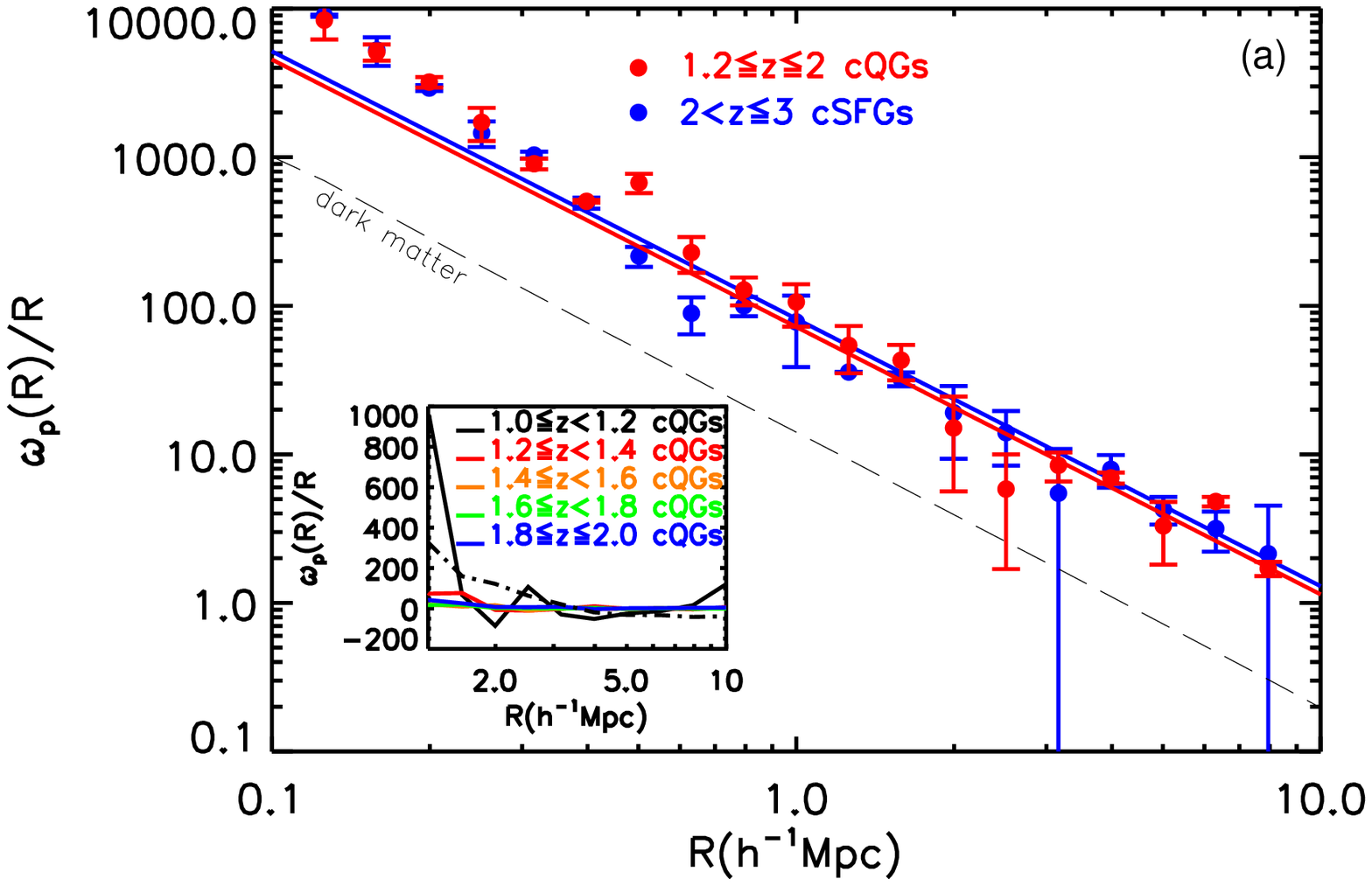}
   \includegraphics[ width=0.495\textwidth]{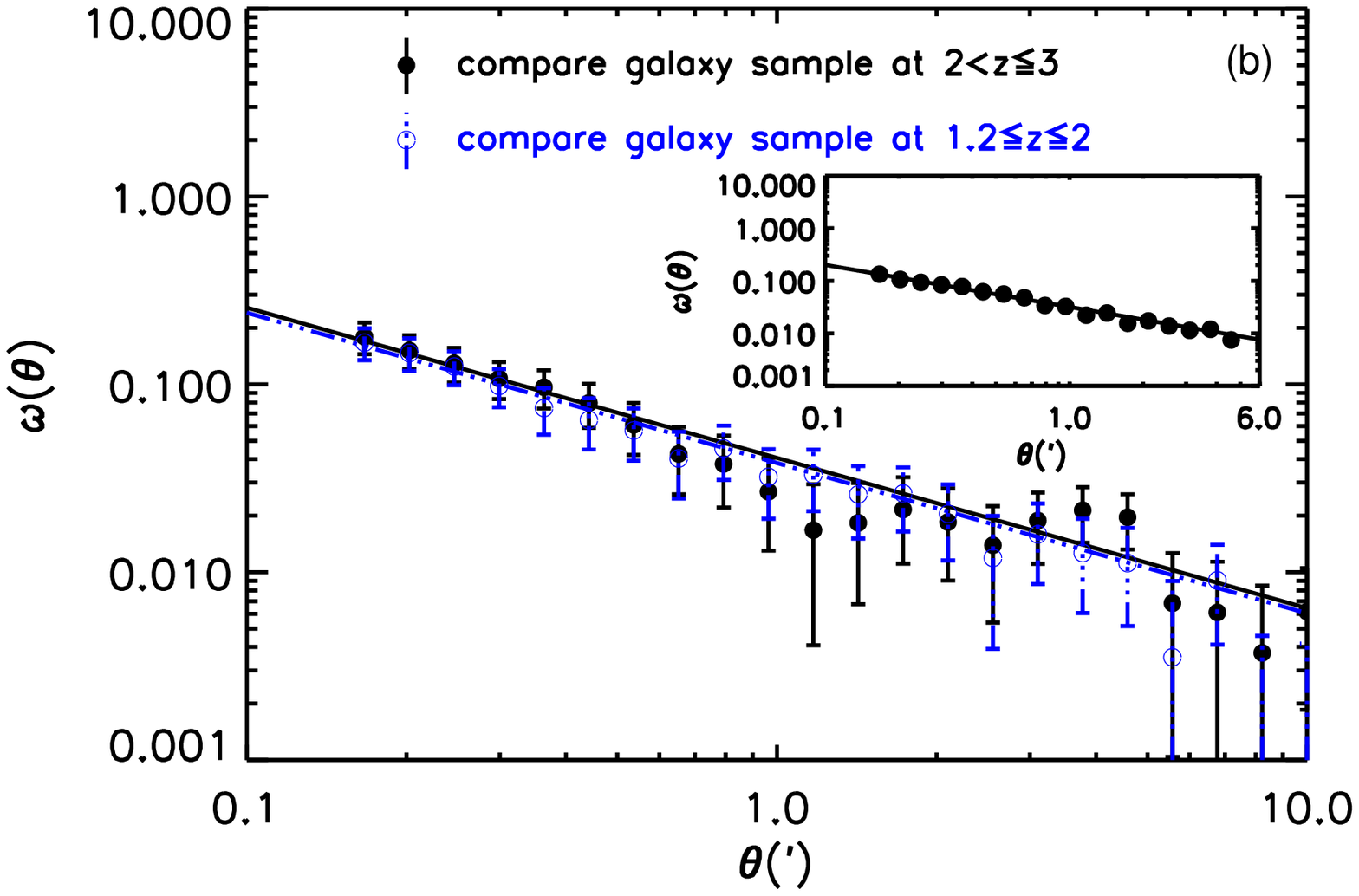}
   \caption{({\em a}) The projected compact and comparison galaxy cross-correlation functions for cQGs at $z\sim1.6$ and cSFGs at $z\sim2.5$. Uncertainties are estimated from bootstrap re-sampling. The solid lines show the best-fit results, adopting the slope $\gamma=1.8$. The dashed line represents the projected correlation function of dark matter. The inset plot shows the projected cross-correlation functions of five cQG subsamples at $1\le z \le 2$, without taking integral constraint into account. { The dot-dashed line in the inset plot shows the cross-correlation function for the cQG subsample at $1\le z<1.2$ without removing the mask regions.} ({\em b}) Angular correlation function $\omega(\theta)$ for comparison galaxy samples at $1.2\le z\le 2$ and $2< z\le3$. Error bars are derived by using the analytic covariance matrix discussed in Brown et al. (2008). Fits are performed on scales of $0.1-10$ arcmins adopting a power-law model with $\delta = 0.8$. The inset plot shows the angular correlation function $\omega(\theta)$ of comparison galaxies at $2< z\le3$ without removing the mask regions.}
\end{center}
   \label{Fig3}
   \end{figure*}

\begin{figure}
\includegraphics[width=\columnwidth]{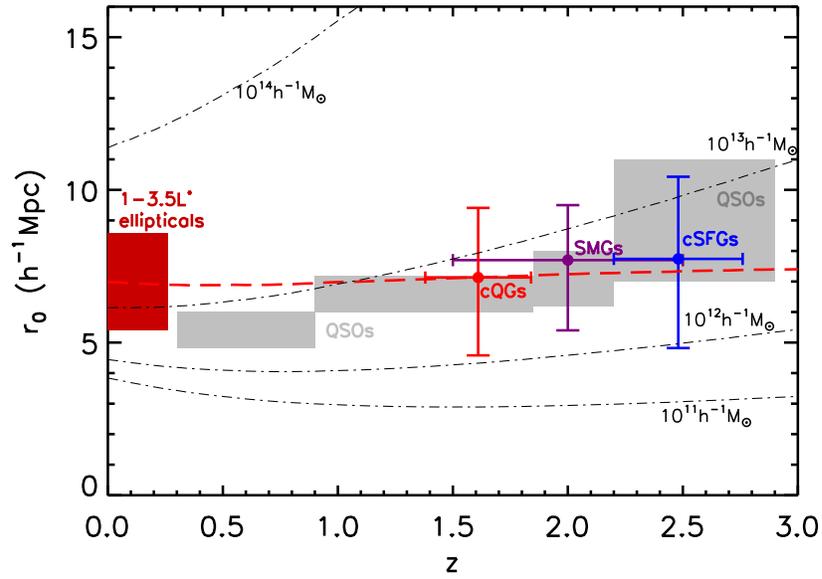}
   \caption{Autocorrelation length $r_{0}$ of different galaxy populations: local luminous early-type galaxies (ETGs) with r-band luminosities of  1.0 to 3.5 $L^{\star}$ (Zehavi et al. 2011), QSOs (Ross et al. 2009; Eftekharzadeh et al. 2015), SMGs at $z\sim2$ (Hickox et al. 2012), cQGs at $z\sim1.6$ and cSFGs at $z\sim2.5$ (this work). Dot-dashed lines show the evolution of $r_{0}$ with redshift for given dark matter halo mass. The dashed line shows the expected evolution in $r_0$ for DM haloes hosting cQGs at $z\sim1.6$, considering the median growth of DM haloes.}
   \label{Fig4}
   \end{figure}

\begin{figure}

   \includegraphics[width=\columnwidth]{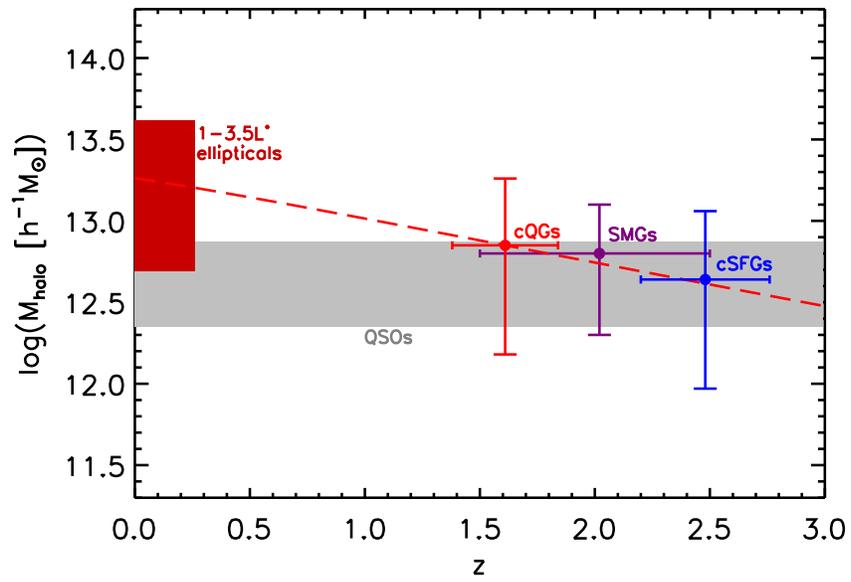}
   \caption{The characteristic halo masses of different galaxy populations derived by the clustering amplitude measurement. The dashed line shows the evolution of halo mass with redshift for cQGs at $z\sim1.6$. }
   \label{Fig5}
\end{figure}

\clearpage

\begin{sidewaystable}[h]
\centering
\caption{Best-fit clustering amplitudes of our galaxy samples in CANDELS/3d-HST\label{tbl-1}}
\begin{tabular}{lccccccccc}
\hline
Sample &
$N_{src}$ &
$\bar{z}$ &
$r_{0}(h^{-1}Mpc)$ &
$b_{C}b_{G}(b_{G}^{2})$ &
$b_{C}(b_{G})$ &
$log (M_H)(h^{-1}M_\odot)$ \\
\hline
cQGs at $1.2\le z\le2$ & 694 & 1.61 & $7.1_{-2.6}^{+2.3}$ & $4.08_{-1.52}^{+1.82}$ & $2.74_{-0.84}^{+0.86}$ & $12.85_{-0.67}^{+0.41}$   \\
cSFGs at $2< z\le3$ & 277 & 2.48 & $7.7_{-2.9}^{+2.7}$ & $7.78_{-3.03}^{+3.64}$ & $3.72_{-1.19}^{+1.23}$ & $12.64_{-0.67}^{+0.42}$    \\
Comparison Galaxies at $1.2\le z\le2$ & 14010 & 1.62 & $3.8_{-0.5}^{+0.5}$ & $2.22_{-0.50}^{+0.54}$ & $1.49_{-0.18}^{+0.17}$ & $11.60_{-0.37}^{+0.27}$    \\
Comparison Galaxies at $2< z\le3$ & 12857 & 2.47 & $4.1_{-0.9}^{+0.5}$ & $4.37_{-1.55}^{+1.01}$ & $2.09_{-0.41}^{+0.23}$ & $11.57_{-0.54}^{+0.23}$  \\
\hline
\end{tabular}
\end{sidewaystable}


\end{document}